\documentclass[prl,article,twocolumn,preprintnumbers,superscriptaddress]{revtex4}
\usepackage{graphicx,amsfonts,color,comment,amsmath,hyperref,float}
\usepackage{amssymb}	
\usepackage{mathrsfs,amssymb}  
\usepackage{cancel}
\usepackage[normalem]{ulem}

\newcommand{\be}{\begin{equation}}
\newcommand{\ee}{\end{equation}}
\newcommand{\bea}{\begin{eqnarray}}
\newcommand{\eea}{\end{eqnarray}}

\newcommand{\cnb}{C$\nu$B}

\begin{document}
%%%%%%%%%%%%%%%%%%%%%%%%%%%%%%%%%%%%%%%%%%%%%%%%%%%%%%%%%%%  FRONT PAGE

\title{Neutrino Echoes from Multimessenger Transient Sources}

\author{Kohta Murase}
\affiliation{Department of Physics; Department of Astronomy and Astrophysics; Center for Particle and Gravitational Astrophysics, The Pennsylvania State University, University Park, Pennsylvania 16802, USA}
\affiliation{Center for Gravitational Physics, Yukawa Institute for Theoretical Physics, Kyoto, Kyoto 16802, Japan}
\author{Ian M. Shoemaker}
\affiliation{Center for Neutrino Physics, Department of Physics, Virginia Polytechnic Institute and State University, Blacksburg, Virginia 24061, USA}
\affiliation{Department of Physics, University of South Dakota, Vermillion, South Dakota 57069, USA}
\date{March 13 2019}

\begin{abstract}
The detection of the high-energy neutrino event, IceCube-170922A, demonstrated that multimessenger particle astrophysics triggered by neutrino alerts is feasible. We consider time delay signatures caused by secret neutrino interactions with the cosmic neutrino background and dark matter and suggest that these can be used as a novel probe of neutrino interactions beyond the Standard Model (BSM). The tests with BSM-induced neutrino echoes are distinct from existing constraints from the spectral modification and will be enabled by multimessenger observations of bright neutrino transients with future experiments such as IceCube-Gen2, KM3Net, and Hyper-Kamiokande. 
The constraints are complementary to those from accelerator and laboratory experiments and powerful for testing various particle models that explain tensions prevailing in the cosmological data.
\end{abstract}

%\preprint{}
%\keywords{}

%%%%%%%%%%%%%%%%%%%%%%%%%%%%%%%%%%%%%%%%%%%%%%%%%%%%%%%%%%%%%%%%%%%
\maketitle

%%%
%{\bf Introduction.~---~}
The new era of multimessenger astroparticle physics has started thanks to the recent detection of high-energy cosmic neutrinos~\cite{Aartsen:2013bka,Aartsen:2013jdh} and gravitational waves~\cite{Abbott:2016blz,TheLIGOScientific:2017qsa}. The detection of the high-energy neutrino event, IceCube-170922A~\cite{Aartsen2018blazar1}, gave further motivation for ``time domain'' particle astrophysics. 
Although the significance of the association with the flaring blazar TXS 0506+056 is only $\sim3\sigma$, this flaring blazar was observed at various wavelengths~\cite{Aartsen2018blazar1}, including x-rays~\cite{Amon2018} and GeV-TeV $\gamma$ rays~\cite{Ahnen:2018mvi}, which demonstrated the capability of multimessenger observations initiated by high-energy neutrino observations. 

Neutrinos have important clues to particle physics Beyond the Standard Model (BSM), as well as the asymmetry between matter and antimatter. Since the discovery of high-energy cosmic neutrinos in IceCube, not only the properties of neutrinos but also different kinds of BSM physics, including dark matter (DM) and nonstandard interactions, have been discussed (see, e.g., \cite{Ahlers:2018mkf,Ackermann:2019cxh}). In the Standard Model (with a minimal extension for finite neutrino masses), the time delay due to the finite neutrino mass ($m_\nu$) is estimated to be $\Delta t\approx m_\nu^2D/(2E_\nu^2)\simeq1.5\times{10}^{-13}~{\rm s}~{(m_\nu/0.1~\rm eV)}^2{(0.1~{\rm PeV}/E_\nu)}^{2}{(D/3~{\rm Gpc})}$, which is much shorter than durations of known astrophysical transients.
% (where $D$ is the source distance).    
Possible time delay between neutrinos and $\gamma$ rays have been discussed to place constraints on the weak equivalence principle (WEP) and Lorentz invariance violation (LIV)~\cite{Wang:2016lne,Ellis:2018ogq,Boran:2018ypz,Laha:2018hsh,Wei:2018ajw}. A time delay of a few days was also reported for IceCube-160731 coincident with a possible $\gamma$-ray counterpart, AGL J1418+0008~\cite{Lucarelli:2017hhh}.  

Not only blazar flares but also various transients, such as long and short $\gamma$-ray bursts (GRBs)~\cite{Paczynski:1994uv,Waxman:1997ti}, supernovae (SNe)~\cite{Murase:2010cu,Murase:2017pfe}, transrelativistic SNe~\cite{Murase:2006mm,Gupta:2006jm}, and tidal disruption events (TDEs)~\cite{Murase:2008zzc,Wang:2011ip}, are promising high-energy neutrino emitters. It is natural that electrons and ions are coaccelerated in these sources, and the temporal and spatial coincidence between neutrinos and $\gamma$ rays is expected. 
Relevant characteristics of various extragalactic transient sources considered in the literature are summarized in Table~1 (see also Refs.~\cite{Guepin:2017dfi,Murase:2019tjj}).   

\begin{figure}[t!]
\includegraphics[angle=0,width=0.5\textwidth]{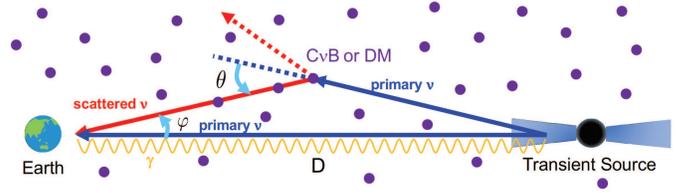}
\caption{Schematic picture of neutrino ``echoes'' induced by BSM interactions. See text for details. }
\label{fig:echo}
\end{figure}

We explore delayed neutrino signatures induced by BSM interactions (see Fig.~\ref{fig:echo}) and suggest that they serve as new probes of secluded interactions with neutrinos themselves and DM particles.  
Probing the parameter space of new interactions is generally important because they cannot be easily probed by terrestrial experiments (e.g.,~\cite{Scholberg:2005qs,Denton:2018xmq,Farzan:2015hkd,Ballett:2019xoj}). 
While our proposed method is applicable to different BSM interactions, we focus on two highly motivated examples that are of broad interest in cosmology and particle physics. It has been shown that models with a light mediator (1) can address all small-scale structure problems~\cite{Aarssen:2012fx,Cherry:2014xra,Loeb:2010gj,Tulin:2012wi,Kaplinghat:2015aga,Tulin:2017ara}, (2) can alleviate Hubble parameter tension~\cite{Riess:2019cxk,Cyr-Racine:2013jua,Escudero:2018mvt,Escudero:2019gzq,Kreisch:2019yzn,Barenboim:2019tux,Park:2019ibn,Forastieri:2019cuf,Blinov:2019gcj}, and (3) may account for the muon anomalous magnetic moment~\cite{Pospelov:2008zw,Altmannshofer:2014pba,Araki:2014ona,Araki:2015mya,Altmannshofer:2016brv,Kamada:2018zxi} and (4) neutrino mass~\cite{Chikashige:1980ui,Gelmini:1980re,Blum:2014ewa}. 
Detections of multiple neutrinos from a transient are feasible with next-generation detectors~\cite{Ackermann:2019ows}. 
BSM tests with neutrino echoes are promising in the upcoming era of time domain multimessenger astrophysics.

 %%%%%%%%%%%%%%%%%%%%%%%%%%%%%%%%%%%%%%%%%%%%%%%%%%
\begin{table}[bt]
\caption{List of extragalactic high-energy neutrino sources, where $\tilde{\mathcal E}_{\rm cr}^{\rm iso}$ is the cosmic-ray energy per logarithmic energy, $D_{{\mathcal N}_{\nu}=1}^{\rm maxeff}$ is the critical distance at which the number of neutrinos detected in IceCube-Gen2~\cite{Aartsen:2014njl} is unity (with the assumption of the maximum neutrino production efficiency), $pp/p\gamma$ is the typical neutrino production channel, ${\Delta T}_{\rm em}$ is the duration of multimessenger emission, and $\rho_0^{\rm em}$ is the local rate density. All values remain as order of magnitude estimates.
%%See Refs.~\cite{Waxman:1997ti,Murase:2005hy,Murase:2008sp,Wang:2008zm,Li:2011ah,He:2012tq,Hummer:2011ms} (LGRB),  Refs.~\cite{Kimura:2017kan,Kimura:2018vvz,Biehl:2017qen} (SGRB), Refs.~\cite{Meszaros:2001ms,Razzaque:2004yv,Ando:2005xi,Murase:2013ffa} (choked jet), Refs.~\cite{Murase:2009pg,Fang:2015xhg,Fang:2018hjp} (SN pulsar), Refs.~\cite{Murase:2010cu,Katz:2011zx,Petropoulou:2017ymv,Murase:2018okz} (SN IIn), Refs.~\cite{Murase:2008zzc,Wang:2011ip,Dai:2016gtz,Senno:2016bso,Lunardini:2016xwi} (Jetted TDE), and \cite{Atoyan:2001ey,Halzen:2005pz,Dermer:2012rg,Dermer:2014vaa,Petropoulou:2016ujj,Halzen:2016uaj,Gao:2016uld} (Blazar flare).
}
\label{tb1}
\begin{center}
% \scalebox{0.9}{%
% \footnotesize
\small
\begin{tabular}{|c||c|c|c|c|c|}
%\multicolumn{5}{c}{TABLE I}\\ \hline
\hline 
Name & $\tilde{\mathcal E}_{\rm cr}^{\rm iso}$ & $D_{{\mathcal N}_{\nu}=1}^{\rm maxeff}$ & $pp/p\gamma$ & ${\Delta T}_{\rm em}$ & $\rho_0^{\rm em}$\\
& [erg] & [Mpc]  &  & [s] &  [${\rm Gpc}^{-3}~{\rm yr}^{-1}$]\\
\hline 
LGRB\footnote{Long $\gamma$-ray bursts. See Refs.~\cite{Waxman:1997ti,Murase:2005hy,Murase:2008sp,Wang:2008zm,Li:2011ah,He:2012tq,Hummer:2011ms}.} 
& ${10}^{52.5}$ & $3000$ & $p\gamma$ & ${10}^{1-2}$ & $0.1-1$\\
\hline
SGRB\footnote{Short $\gamma$-ray bursts. See Refs.~\cite{Kimura:2017kan,Kimura:2018vvz,Biehl:2017qen}.} 
& ${10}^{50.5}$ & $300$ & $p\gamma$ & $0.1-1$ & $10-100$\\
\hline
SN (choked jet)\footnote{Supernovae powered by choked jets. See Refs.~\cite{Meszaros:2001ms,Razzaque:2004yv,Ando:2005xi,Murase:2013ffa}.} 
& ${10}^{50.5}$ & $300$ & $p\gamma$ & ${10}^{1-4}$ & $10^2-10^3$ \\
\hline
SN (pulsar)\footnote{Supernovae powered by pulsar winds. See Refs.~\cite{Murase:2009pg,Fang:2015xhg,Fang:2018hjp}.} 
& ${10}^{50}$ & $200$ & $pp$& ${10}^{3-6}$ & $10^{3.5}-10^{4.5}$ \\
\hline
SN (IIn)\footnote{Type IIn supernovae powered by shocks. See Refs.~\cite{Murase:2010cu,Katz:2011zx,Petropoulou:2017ymv,Murase:2018okz}.} 
& ${10}^{49}$ & $50$ & $pp$ & ${10}^{6-7}$  &$10^4$ \\
\hline
Jetted TDE\footnote{Jetted tidal disruption events. See Refs.~\cite{Murase:2008zzc,Wang:2011ip,Dai:2016gtz,Senno:2016bso,Lunardini:2016xwi}.} 
& $10^{53}$ & $5000$ & $p\gamma$ & $10^{6-7}$ & $0.01-0.1$\\
\hline
Blazar flare\footnote{See Refs.~\cite{Atoyan:2001ey,Halzen:2005pz,Dermer:2012rg,Dermer:2014vaa,Petropoulou:2016ujj,Halzen:2016uaj,Gao:2016uld}.} 
& ${10}^{54}$ & $15000$ & $p\gamma$& ${10}^{5-7}$ & $0.1-1$\\
\hline
\end{tabular}
% }
\end{center}
\vspace{0.5\baselineskip}
%\caption{List of extragalactic high-energy neutrino sources, where $\tilde{\mathcal E}_{\rm cr}^{\rm iso}$ is the cosmic-ray energy per logarithmic energy, $D_{{\mathcal N}_{\nu=1}}^{\rm maxeff}$ is the critical distance at which the number of neutrinos detected in IceCube-Gen2~\cite{Aartsen:2014njl} is unity (with the assumption of the maximum neutrino production efficiency), $pp/p\gamma$ is the typical neutrino production channel, ${\Delta T}_{\rm em}$ is the typical duration of electromagnetic emission, and $\rho_0^{\rm em}$ is the local rate density of the sources. Note that all values remain as order of magnitude estimates. 
%See Refs.~\cite{Waxman:1997ti,Murase:2005hy,Murase:2008sp,Wang:2008zm,Li:2011ah,He:2012tq,Hummer:2011ms} (LGRB),  Refs.~\cite{Kimura:2017kan,Kimura:2018vvz,Biehl:2017qen} (SGRB), Refs.~\cite{Meszaros:2001ms,Razzaque:2004yv,Ando:2005xi,Murase:2013ffa} (choked jet), Refs.~\cite{Murase:2009pg,Fang:2015xhg,Fang:2018hjp} (SN pulsar), Refs.~\cite{Murase:2010cu,Katz:2011zx,Petropoulou:2017ymv,Murase:2018okz} (SN IIn), Refs.~\cite{Murase:2008zzc,Wang:2011ip,Dai:2016gtz,Senno:2016bso,Lunardini:2016xwi} (Jetted TDE), and \cite{Atoyan:2001ey,Halzen:2005pz,Dermer:2012rg,Dermer:2014vaa,Petropoulou:2016ujj,Halzen:2016uaj,Gao:2016uld} (Blazar flare).}
\end{table}
%%%%%%%%%%%%%%%%%%%%%%%%%%%%%%%%%%%%%%%%%%%%%%%%%%

{\bf Example 1: Neutrino self-interactions.~---~}
It has been discussed that BSM-induced neutrino self-interactions may occur~\cite{BialynickaBirula:1964zz,Bardin:1970wq}, and models can generate finite neutrino masses~\cite{Chikashige:1980ui,Gelmini:1980re,Blum:2014ewa}. We consider such nonstandard, secret neutrino interactions that may lead to effective Lagrangians, e.g., ${\mathcal L}\supset g_{ij}\bar{\nu}_i\nu_j\phi$ (for scalars), ${\mathcal L}\supset g_{ij}\bar{\nu}_i(i\gamma^5\phi)\nu_j$ (for pseudoscalars), and ${\mathcal L}\supset g_{ij}\bar{\nu}_i(\gamma^\mu V_\mu)\nu_j$ (for vector bosons), where $g_{ij}$ is the coupling parameter. Note that, although we do not specify whether neutrinos are Dirac or Majorana types, the allowed interactions for scalars and pseudoscalars are, e.g., ${\mathcal L}\supset g\nu_L\nu_L\phi+{\rm c.c.}$ and ${\mathcal L}\supset gN_RN_R\phi+{\rm c.c.}$, where $\nu_L$ is the left-handed neutrino and $N_R$ is the right-handed neutrino. 
Remarkably, it has been shown that a $1-100$~MeV scale mediator also enables us to resolve various cosmological issues such as the tension in the Hubble parameter~\cite{Escudero:2019gzq,Kreisch:2019yzn,Barenboim:2019tux} and the missing satellite and core-cusp problems~\cite{Aarssen:2012fx,Cherry:2014xra}.
With the mediator mass $m_\phi$, the resonance interaction with the cosmic neutrino background (C$\nu$B) happens at $E_\nu=m_\phi^2/(2m_\nu)\simeq1.25\times{10}^{14}~{\rm eV}~{(m_\phi/5~{\rm MeV})}^2{(m_\nu/0.1~{\rm eV})}^{-1}$, corresponding to the IceCube energy range~\cite{Ioka:2014kca,Ng:2014pca,Ibe:2014pja,Araki:2014ona,Blum:2014ewa,Cherry:2014xra,DiFranzo:2015qea,Araki:2015mya,Shoemaker:2015qul,Kelly:2018tyg,Barenboim:2019tux}. 

%The differential cross section for the two-body scattering $A+B\rightarrow C+D$ in the rest frame of $B$ is given by:
%\be 
%\left(\frac{d\sigma}{d\Omega}\right)=\frac{1}{64 \pi^{2}p_Am_A} \frac{{|p_{C}|}^2}{|p_{C}|(E_A+m_B)-E_C|p_A|\cos\theta} |\mathcal{M}|^{2},
%\ee
%where $|\mathcal{M}|^{2}$ is the spin-averaged matrix elements. 
%Then, the average scattering angle is evaluated via:
%\be
%\langle(1-\cos\theta)\rangle=\frac{1}{\sigma}\int d\Omega \, (1-\cos\theta) \left(\frac{d\sigma}{d\Omega}\right).
%\ee 
Let us consider the neutrino-(anti)neutrino scattering process via $s$ channel, $\nu\nu\rightarrow\phi\rightarrow\nu\nu$. 
In this case, the angular distribution of the scattered neutrinos is isotopic in the center-of-momentum frame. (In general, details depend on the mediator spin as well as the main scattering channel.) In the \cnb\ frame, because of the boost $\sim E_\nu/\sqrt{s}\sim\sqrt{E_\nu/m_\nu}$, we may write
\begin{equation}
\sqrt{\langle\theta^2\rangle}\approx C\frac{\sqrt{s}}{E_\nu}\simeq4.5\times{10}^{-8}~C{\left(\frac{m_\nu}{0.1~{\rm eV}}\right)}^{\frac{1}{2}}{\left(\frac{0.1~{\rm PeV}}{E_\nu}\right)}^{\frac{1}{2}},
\end{equation}
where $\theta$ is the scattering angle and $C\sim1$ for a scalar or pseudoscalar mediator in the neutrino-neutrino scattering.  
%$C=\sqrt{[1+{(2\omega)}^{-1}]\ln(1+2\omega)-1}$ for a scalar or pseudoscalar mediator, which is obtained in the limit of $E_\nu\gg m_\nu$, and $\omega=E_\nu/m_\nu$. 
More generally, for the differential cross section ($d\sigma/d\Omega$),  the average scattering angle is evaluated via 
\be
\langle(1-\cos\theta)\rangle=\frac{1}{\sigma}\int d\Omega \, (1-\cos\theta) \left(\frac{d\sigma}{d\Omega}\right).
\ee 
For example, $E_\nu=0.1$~PeV and $m_\nu=0.1$~eV leads to $\langle\theta\rangle\approx2.8\times{10}^{-8}$ for a leading neutrino. 
Resulting angular spreading may be too small to be seen as a ``halo'' around the source, but can be big enough to make a sizable time delay signal (``neutrino echo"). The geometrical setup is analogous to $\gamma$-ray ``pair echoes'' proposed as a probe of intergalactic magnetic fields~\cite{Plaga:1995ins,Ichiki:2007nd,Murase:2008pe,Takahashi:2008pc,Murase:2008yy,Murase:2009ah}, although underlying interaction processes are completely different. 
Neutrinos scattering during propagation was discussed for SN 1987A~\cite{Kolb:1987qy,Lindner:2001th}, but detailed methodology to utilize the time delay has not been studied.

{\it Large optical depth (conservative) limit:}
So far, the expected number of high-energy neutrinos is limited. However, even if statistics are not large, e.g., ${\mathcal N}_{\nu}\sim{\rm a~few}$, the sizable effect of BSM interactions exists if the optical depth to the neutrino scattering is larger than unity,
\be
\tau_{\nu}=n_{\nu}\sigma_{\nu}D\gtrsim1.
\ee
The probability for neutrinos to experience the neutrino scattering is given by $1-\exp(-\tau_\nu)$. In the large $\tau_\nu$ limit, most of the neutrinos are scattered, and the spectral and flux information can be used to probe BSM neutrino interactions~\cite{Shoemaker:2015qul,Arguelles:2017atb,Kelly:2018tyg}. Large statistics would also be required, and the current constraints are much weaker than the ideal bound placed by $n_\nu\sigma_{\nu}H_0<1$ (where $H_0$ is the Hubble constant). 
Although the diffuse neutrino limits can be relevant, Ref.~\cite{Shoemaker:2015qul} showed that such an ideal limit [e.g., $g\lesssim3\times{10}^{-4}~(m_\phi/10~{\rm MeV})$ in the scalar mediator case] can be achieved for $m_\phi\sim20-30$~MeV with ten years of observations by IceCube-Gen2. 
As we see below, the time delay argument can provide us with a meaningful limit even with limited statistics, without relying much on the spectral information.

In the multiple scattering case, neutrino cascades~\cite{Ioka:2014kca,Ng:2014pca} occur and the arrival angle averaged over scatterings is given by $\langle\varphi^2\rangle\approx(\tau_{\nu}/3)\langle\theta^2\rangle\propto n_\nu \sigma_{\nu} D E_\nu^{-1}$. 
The corresponding characteristic time delay is
\begin{eqnarray}
{\Delta t}\approx\frac{1}{4}\langle\varphi^2\rangle D&\simeq&500~{\rm s}~\left(\frac{\tau_\nu}{10}\right)\left(\frac{D}{3~\rm Gpc}\right)\nonumber\\
&\times&C^2{\left(\frac{m_\nu}{0.1~{\rm eV}}\right)}{\left(\frac{0.1~{\rm PeV}}{E_\nu}\right)}.
\label{delay1}
\end{eqnarray}

If the neutrinos arrive within a time window of $\Delta T$ that may be the duration of intrinsic multimessenger emission (${\Delta T}_{\rm em}$), possible constraints can be placed by $\Delta t<\Delta T$, which leads to
\be
\sigma_{\nu}\lesssim\frac{12\Delta T}{D^2n_\nu\langle\theta^2\rangle}.
\label{constraint1}
\ee
This is valid only if $D\langle\theta^2\rangle\lesssim8\Delta T$, otherwise the time delay itself does not give a direct constraint on the cross section because of $\tau_\nu\lesssim1.5$. In the neutrino-neutrino scattering case, this implies $\Delta T \gtrsim 30~{\rm s}~C^2(D/1~{\rm Gpc})(m_\nu/0.1~{\rm eV}){(E_\nu/0.1~{\rm PeV})}^{-1}$.
The detection of neutrinos with $E_\nu$ implies that some neutrinos arrive without significant energy losses, for which Eq.~(\ref{constraint1}) is applied~\footnote{The flux of survived neutrinos that do not lose significant energies by multiple scatterings is suppressed for $\tau_\nu\gtrsim{\mathcal M}$.
%The flux of survived neutrinos, for which energy losses are small enough, is $F_{E_\nu}^{\rm sur}(\Delta t; D)\approx[4\pi^2 \phi_\nu^{\rm sur}/(3\langle\varphi^2‘\rangle D)]\Sigma_{n=1}^{\infty}{(-1)}^{n+1}n^2\exp[-2n^2\pi^2\Delta t/(3\langle\varphi^2\rangle)]$, where $\phi_\nu^{\rm sur}$ is the fluence of survived neutrinos (see Ref.~\cite{1978ApJ...222..456A} for the formulation).
}. 
If one requires the bulk of neutrinos with $E_\nu$ survives after $\mathcal M$ scatterings, an additional constraint, $\tau_\nu\lesssim{\mathcal M}$, may be imposed, but the actual limits depend on the unknown primary fluence and spectrum. Eq.~(\ref{constraint1}) typically leads to conservative limits.
Note that for $\tau_\nu\gg1$ most neutrinos are cascaded down and appear at sufficiently lower energies. If the optical depth for the cascaded component is less than unity, the bulk of the delayed flux is roughly estimated by $F_{E_\nu}^{\rm cas}(t)\sim\int d\tilde{\theta}\, 4{[2\pi\langle{\tilde{\varphi}}^2(t,\tilde{\theta})\rangle]}^{-1/2}{[{\tilde{\theta}}^2+\langle{\tilde{\varphi}}^2(t,\tilde{\theta})\rangle]}^{-1}e^{-{\tilde{\theta}}^2/[2\langle{\tilde{\varphi}}^2(t,\tilde{\theta})\rangle]}$\\$F_{E_\nu}^{\rm cas0}$, where $F_{E_\nu}^{\rm cas0}$ is the flux of cascaded neutrinos in the absence of angular spreading~\cite{Ichiki:2007nd}. The characteristic time delay of this cascaded component is estimated to be $\Delta t_{\rm cas}\sim(1/12)\langle\theta^2\rangle{\mathcal M}/(n_\nu\sigma_\nu)$ (cf.~Eq.~\ref{delay1}). The full radiative transfer calculation is necessary to consistently describe the echo flux for arbitrary $E_\nu$ and $\tau_\nu$.

{\it Small optical depth (stronger) limit.---}
The constraints discussed above make sense when the coupling is so large that multiple scattering events occur. However, this may not be possible for several reasons. 
First, the coupling or the scattering cross section may be bounded by other existing constraints, so that $\sigma_\nu$ cannot be large enough. 
Second, the condition $D\langle\theta^2\rangle\lesssim8\Delta T$ is not satisfied. For example, $\tau_\nu\gtrsim1-2$ is prohibited if the observed time window $\Delta T$ is too short. 
On the other hand, bright neutrino transients such as choked GRB jets and blazar flares could be detected with a large number of signals (i.e., ${\mathcal N}_\nu\gg1$) by future neutrino telescopes such as IceCube-Gen2 and KM3Net, in which we may still obtain useful constraints that can actually be better than those from Eq.~(\ref{constraint1}) and even exceed the mean free path limit~\cite{Shoemaker:2015qul,Kelly:2018tyg}.

In the small $\tau_\nu$ limit, most of neutrinos ($\sim {\mathcal N}_\nu$) are expected to arrive together with photons within $\Delta T={\Delta T}_{\rm em}$. However, in the presence of the BSM neutrino scattering, some neutrinos ($\sim \tau_\nu {\mathcal N}_\nu$) experience the scattering once during the propagation, and the characteristic time delay is given by:
\begin{equation}
{\Delta t}\approx\frac{1}{2}\frac{\langle\theta^2\rangle}{4} D\simeq77~{\rm s}~\left(\frac{D}{3~\rm Gpc}\right)C^2{\left(\frac{m_\nu}{0.1~{\rm eV}}\right)}{\left(\frac{0.1~{\rm PeV}}{E_\nu}\right)}.
\label{delay2}
\end{equation}
This expression does not include $\sigma_\nu$, and with Eq.~(\ref{delay1}) the time delay is estimated by $\Delta t\approx{\rm max}[\langle\varphi^2\rangle D/4,\langle\theta^2\rangle D/8]$.
The probability distribution of delayed neutrinos in the small $\tau_\nu$ limit is expressed as $P(t,\varphi; D)\approx1/[t+(D\varphi^2/2)](1/\sigma_\nu)(d\sigma_\nu/d\theta)|_{\theta=\varphi+2t/(D\varphi)}$~\cite{1978ApJ...222..456A}. We remark that only one scattering matters and the time delay distribution reflects the differential cross section of the neutrino-neutrino scattering that is generally inelastic. 

Given ${\mathcal N}_{\nu}\gg1$, stronger limits can be placed for $\Delta T\lesssim\langle\theta^2\rangle D/8$ (implying $\tau_\nu\lesssim1.5$), in which nondetection of time delayed events itself may be used.
In the limit that the atmospheric background is negligible, 
the sizable effect is observable when the number of delayed signals is larger than unity, i.e., $\tau_\nu\gtrsim1/{\mathcal N}_\nu$. If the background is not negligible, one would need $\tau_\nu\gtrsim\sqrt{{\mathcal N}_{\nu}^{\rm bkg}}/{\mathcal N}_\nu$, where ${\mathcal N}_{\nu}^{\rm bkg}$ is the number of background events for a given time window.  
In the background-free regime (that is valid for short duration transients), nondetection of echoes gives:
\be
\sigma_{\nu}\lesssim\frac{2.3}{{\mathcal N}_\nu n_\nu D},
\label{constraint2}
\ee
where the Poisson probability to observe nonzero time delayed events is set to $<0.9$. 
One should keep in mind that the neutrino scattering cross section is energy dependent and $D\langle\theta^2\rangle\gtrsim8\Delta T$ should be satisfied. Note that Eq.~(\ref{constraint1}) is applied in the opposite limit. 

\begin{figure}[t!]
\includegraphics[angle=0,width=0.5\textwidth]{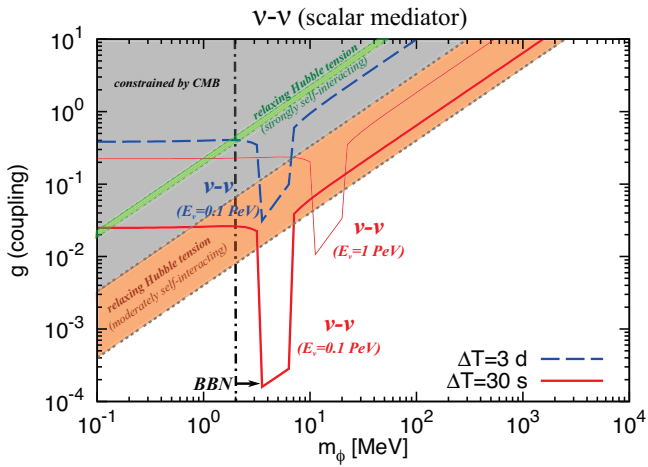}
\caption{Expected neutrino echo constraints on secret neutrino interactions via a scalar mediator. The distance and neutrino mass are $D=3$~Gpc and $m_\nu=0.1$~eV, respectively, and ${\mathcal N}_\nu=10$ is used for the small optical depth limit. 
The parameter space relaxing the Hubble parameter tension for the cosmic microwave background (CMB)~\cite{Kreisch:2019yzn,Blinov:2019gcj} is shown together with constraints assuming $\Lambda$CDM cosmology (shaded regions).} 
\label{fig:constrain1}
\end{figure}

We show results for a scalar mediator in Fig.~\ref{fig:constrain1}. Here contributions from $t$ and $u$ channels are also included~\cite{Ioka:2014kca,Blum:2014ewa}. In the resonant region ($s\sim m_\phi^2$), we average the effective cross section by assuming an energy resolution of $\Delta \log(E_\nu)=0.6$ (which is reasonable for high-energy track events~\cite{Ackermann:2019ows}). At $E_\nu=0.1$~PeV, the two cases of $\Delta T=3$~d and $\Delta T=30$~s correspond to the large and small optical depth limits, respectively. We also show another case of $\Delta T=30$~s for $E_\nu=1$~PeV, in which the multiple scattering limit is applied. 

Other constraints include one from kaon decay, which gives $g\lesssim0.01$~\cite{Blum:2014ewa,Lessa:2007up,Laha:2013xua}. 
Note that our echo method is especially relevant if only tau neutrinos have BSM interactions.
Big Bang Nucleosynthesis (BBN) gives a constraint of $m_\phi\gtrsim{\rm a~few}$~MeV, although details depend on uncertainty in the extra number of relativistic species (e.g.,~\cite{Aarssen:2012fx,Ahlgren:2013wba,Blinov:2019gcj}). 
Astrophysical and laboratory limits are complementary. 
For example, if neutrinos interact with the \cnb\, through sterile neutrinos, the limits can be relaxed, depending on mixing angles~\cite{Cherry:2014xra,Shoemaker:2015qul}.

\begin{figure}[t!]
\includegraphics[angle=0,width=0.5\textwidth]{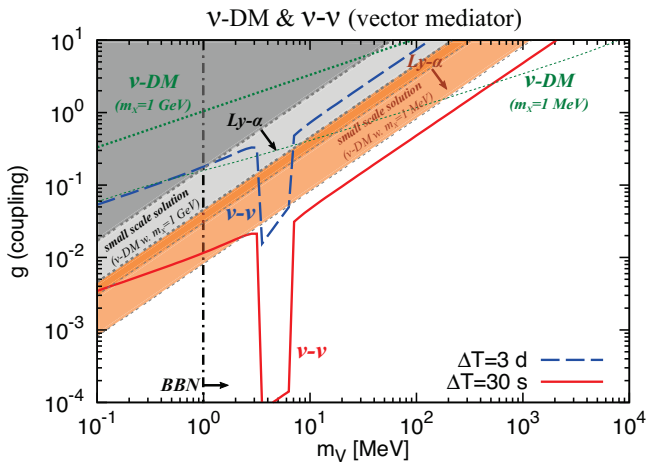}
\caption{Expected constraints on secret neutrino interactions via a vector mediator in the presence of DM. The neutrino energy is set to $E_\nu=0.1$~PeV, and $D$, $m_\nu$ and ${\mathcal N}_\nu$ are the same as in Fig.~\ref{fig:constrain1}. Lyman-$\alpha$ constraints from the kinetic decoupling for neutrino-DM scatterings are shown as conservative limits for different DM masses. The parameter space proposed to solve the small-scale structure abundance problem~\cite{Aarssen:2012fx} is also indicated (light shaded regions). 
The CMB constraints shown in Fig.~\ref{fig:constrain1} are applied to the neutrino-neutrino scattering.} 
\label{fig:constrain2}
\end{figure}

{\bf Example 2: Neutrino-DM interactions.~---~}
As a further application of the idea of BSM-induced neutrino echoes, we discuss neutrinophilic DM models in which DM and neutrinos share a new interaction. 
Very intriguingly, such models give a possible solution to cosmological issues~\cite{Aarssen:2012fx,Cherry:2014xra,Tulin:2017ara,Escudero:2018mvt,Escudero:2019gzq,Park:2019ibn} and can explain the muon anomalous magnetic moment~\cite{Araki:2014ona,Araki:2015mya,Kamada:2018zxi}.
For illustration, we consider a simple extension of the vector model mentioned above in which the new gauge boson also couples to a Dirac fermion DM, $\mathcal{L}\supset g_\nu V_{\mu}\bar{\nu}\gamma^{\mu}\nu+g_XV_{\mu}\bar{X}\gamma^{\mu}X$, where $X$ denotes the DM with a mass $m_{X}$. 
New gauge bosons appear in many BSM scenarios~\cite{Langacker:2008yv}, and additional broken U(1) gauge symmetries leading to vector bosons were predicted by grand unification theories~\cite{London:1986dk,Hewett:1988xc}.  
While the neutrinos and DM may have different charge assignments, here we take them to be equal and assume $g_\nu=g_X=g$.

The above model is accompanied by neutrino-DM scatterings, and the resulting constraints are shown in Fig.~\ref{fig:constrain2}. 
As in the previous case, if a bright neutrino transient with short duration is observed, we may place strong constraints even in the small optical depth limit, which can be more stringent than previous ones~\cite{deSalas:2016svi,Reynoso:2016hjr,Arguelles:2017atb,Kelly:2018tyg,Pandey:2018wvh,Choi:2019ixb}.
Here the coupling should be regarded as an effective parameter. The real coupling to the Standard Model can be made neutrinophilic via coupling the gauge boson to heavy sterile neutrinos.  However, their effect is still felt as they effectively endow the active neutrinos with a mixing suppressed coupling to the new mediator. Such models have been explored in Refs.~\cite{Pospelov:2011ha,Bertoni:2014mva,Batell:2017cmf}.  

For the $t$ channel, we find that the multiple scattering limit may not be applicable to most transients due to large values of $\langle\theta^2\rangle$ for relatively heavy DM. 
The cases for $\Delta T=30$~s are shown in Fig.~\ref{fig:constrain2}, where the constraint is given for the small optical depth limit (but with the replacement of $n_\nu$ with $n_X$). 
As we see, the limits are more stringent for lower-mass mediators. The resulting constraint is comparable to that expected from detailed analyses with spatial and spectral information~\cite{Arguelles:2017atb}.  

We note that the time delay from neutrino-DM scatterings receives contributions from both the Milky Way DM halo and extragalactic DM components. As known for decaying DM signals, the DM located in the line of sight is almost comparable because of $R_{\rm MW}\varrho_{X}^{\rm local}\sim H_0^{-1}\varrho_{X}\gtrsim D\varrho_{X}$, where $R_{\rm MW}\sim10$~kpc is the typical size of the Milky Way. 
For the Galactic contribution, the condition $\Delta T\gtrsim R_{\rm MW}\langle\theta^2\rangle/8$ is more easily satisfied, which may lead to $\sigma_{\nu X}\lesssim5.4\times{10}^{-24}~{\rm cm}^2~{(\Delta T/1~{\rm d})}{(R_{\rm MW}/10~\rm kpc)}^{-2}C^{-2}~{(E_\nu/0.1~{\rm PeV})}$. Here $C$ depends on $\sim m_V/\sqrt{s}$ for the $t$ channel.
For models that lead to sufficiently small scattering angles, the time delay in the large optical depth limit becomes independent of the DM mass, implying $\sigma_{\nu X} \lesssim10^{-28}~{\rm cm}^{2}~{(\Delta T/1~{\rm d})}{(D/1~\rm Gpc)}^{-2}C^{-2}~{(E_\nu/0.1~{\rm PeV})}$. Although such limits would be weaker than the cosmology limits, $\sigma_{\nu X} \lesssim 10^{-33}~{\rm cm}^{2}$~\cite{Wilkinson:2014ksa}, it takes place at much higher center-of-momentum energies.

Finally, we comment on other constraints that can be relevant. If neutrino-DM scatterings are efficient in the early Universe it can inject energy and potentially ``heat'' the cold DM such that Lyman-$\alpha$ bounds on the small-scale structure are violated~\cite{Boehm:2000gq,Hooper:2007tu,Aarssen:2012fx,Wilkinson:2014ksa}. This effect can be used to explain small-scale structure problems of cold DM~\cite{Aarssen:2012fx}, and the region favored by this argument is shown in Fig.~\ref{fig:constrain2}. Couplings above these regions are excluded. Additionally, note that neutrinophilic DM should not thermalize for DM masses at the MeV scale~\cite{Boehm:2013jpa}, although a narrow window of thermal neutrinophilic DM exists below a MeV~\cite{Berlin:2017ftj,Berlin:2018ztp}. Finally, in models with direct couplings to active neutrinos laboratory constraints from $Z$ and meson decays can be strong~\cite{Lessa:2007up,Laha:2013xua,Araki:2015mya}.

%%%
%\section{Summary and Discussion}
{\bf Summary and discussion.---}
We proposed detailed time delay signatures as a novel probe of BSM neutrino interactions. 
Notably, BSM-induced neutrino echoes generally predict $\Delta t\propto E_\nu^{-1}C^2$. This is distinct from predictions of other BSM signatures such as LIV and WEP violation (see a review \cite{Ahlers:2018mkf}). 
For example, LIV shifts the light velocity by ${(E_\nu/\zeta_n M_{\rm pl})}^n$ (where $M_{\rm pl}$ is the Planck mass), leading to $\Delta t=D(E_\nu/\zeta_n M_{\rm pl})^n$ (e.g.,~\cite{Jacob:2006gn,Murase:2009ah}). 
%This implies that the LIV-induced time delay increases with energy.  
%$\frac{(1+n)E_\nu^n}{2E_{\rm QG}^n} \int_{0}^{z}\frac{(1+z')^{n}}{H(z')} dz'$
%$\Delta t = \frac{1+n}{2H_{0}} \frac{E^{n}}{E_{QG}^{n}}\int_{0}^{z}\frac{(1+z')^{n}}{\sqrt{\Omega_{\Lambda} + \Omega_{M}(1+z')^{3}}} dz'$
For neutrino-neutrino scatterings, cosmological time delays are dominant. On the other hand, the Milky Way DM contributes to neutrino-DM scatterings. This implies that DM in the host galaxy may also contribute to the time delay depending on $\Delta T$ and $\langle\theta^2\rangle$. 

Neutrino echo constraints can be placed without large statistics of neutrino events. If a bright transient occurs, we can even go beyond the mean free path limit without relying on details of neutrino spectra. 
Nondetection of echoes lead to powerful constraints as long as we only have coincident detections. Possible detections open up a new window for BSM neutrino physics. In the small optical depth limit that is more likely, we could directly measure the differential cross section. In the other limit, the delayed cascaded component accompanied by the strong absorption feature in the neutrino spectrum serves as a testable prediction. 
The relevant parameter space probed by current and future multimessenger observations of neutrino transients is complementary to those from accelerator and laboratory experiments such as DUNE~\cite{Farzan:2015hkd,Ballett:2019xoj} and COHERENT~\cite{Scholberg:2005qs,Denton:2018xmq}. 
We demonstrated that our method is particularly powerful for light mediator models, which are extensively discussed in the context of ``self-interacting neutrino cosmology''~\cite{Cyr-Racine:2013jua,Escudero:2018mvt,Escudero:2019gzq,Barenboim:2019tux,Park:2019ibn,Wilkinson:2014ksa} and substructure problems~\cite{Aarssen:2012fx,Cherry:2014xra,Tulin:2017ara}, and has been invoked to explain the muon anomalous magnetic moment~\cite{Pospelov:2008zw,Altmannshofer:2014pba,Araki:2014ona,Araki:2015mya,Altmannshofer:2016brv,Kamada:2018zxi} and neutrino masses~\cite{Chikashige:1980ui,Gelmini:1980re,Blum:2014ewa}. 

We provided intriguing examples for ``time domain'' multimessenger astroparticle physics, cosmology, and particle physics. 
BSM tests with neutrino echoes are general and various types of BSM interactions~\cite{Ackermann:2019cxh} including the long-range one~\cite{Bustamante:2018mzu} could be considered. 
Neutrino transients (e.g., supernovae) should exist and the proposed method is applicable to lower-energy (e.g., GeV--TeV) neutrino transients including the next Galactic supernova, for which $\sim100-1000$ events of high-energy neutrinos can be detected in IceCube and KM3Net~\cite{Murase:2017pfe}. 
Even supernova neutrino bursts in the MeV range will give us useful constraints, because the time delay is $\Delta t\simeq260~{\rm s}~{(D/10~{\rm kpc})}C^2{(m_\nu/0.1~{\rm eV})}{(E_\nu/100~{\rm MeV})}^{-1}$. 
Detections are even more promising for next-generation neutrino detectors such as IceCube-Gen2~\cite{Aartsen:2014njl}, KM3Net~\cite{Adrian-Martinez:2016fdl} and Hyper-Kamiokande~\cite{Abe:2018uyc}. Searches can also be performed for not only a single transient but also ``stacked'' samples of short duration transients.

%\begin{acknowledgements}
{\bf Acknowledgements.---}
We thank Peter Denton, Walter Winter, and an anonymous referee for useful comments. 
This work is supported by the Alfred P. Sloan Foundation, NSF Grants No.~PHY-1620777 and No.~AST-1908689 (K.M.), as well as the U.S. Department of Energy under Award No.~DE-SC0020250 (I.M.S.).
K.M. acknowledges the Munich Institute for Astro- and Particle Physics (MIAPP) of the DFG cluster of excellence ``Origin and Structure of the Universe'', and he advanced this project during the workshop in February -- March 2018. 
As the manuscript was being completed, we became aware of a related but independent work~\cite{Koren:2019wwi}. 
%\end{acknowledgements}

%\vspace{2cm}

%\bibliographystyle{JHEP}
\bibliography{kmurase,nu}

\end{document}